\documentclass[aps,prc,groupedaddress,twocolumn,showpacs,amsmath,amssymb]{revtex4}
\usepackage{graphics}
\usepackage{dcolumn}
\usepackage{bm}

\begin{document}

\title{Effects of nuclear symmetry energy and in-medium $NN$ cross section in
    heavy-ion collisions at beam energies below the pion production
    threshold}

\author{Wen-Mei Guo$^{1,2,3}$}
\author{Gao-Chan Yong$^{1,4,5}$}\email{yonggaochan@impcas.ac.cn}
\author{Wei Zuo$^{1,4,5}$}

\affiliation{%
$^1${Institute of Modern Physics, Chinese Academy of Sciences, Lanzhou 730000, China}\\
$^2${School of Physical Science and Technology, Lanzhou University,
Lanzhou 730000, China}\\
$^3${University of Chinese Academy of Sciences, Beijing 100049, China}\\
$^4${State Key Laboratory of Theoretical Physics, Institute of
Theoretical Physics, Chinese Academy of Sciences, Beijing,
100190}\\
$^5${Kavli Institute for Theoretical Physics, Chinese Academy of Sciences, Beijing 100190, China}\\
}%

\date{\today}

\begin{abstract}
Based on the isospin-dependent Boltzmann-Uehling-Uhlenbeck(IBUU04)
transport model, we explored effects of in-medium $NN$ elastic
scattering cross section and nuclear symmetry energy on the
sub-threshold pion production in $^{132}Sn+^{124}Sn$ reaction. We
find that with decrease of incident beam energy, effects of
in-medium $NN$ elastic scattering cross section on the
$\pi^{-}/\pi^{+}$ ratio are larger than that of the symmetry
energy although the latter may be also larger. While keeping the
effect of symmetry energy, the double ratio of $\pi^{-}/\pi^{+}$
from neutron-rich and neutron-poor reaction systems (with the same
mass number of system) $^{132}Sn+^{124}Sn$ and $^{128}Pm+^{128}Pm$
almost fully cancels out the effects of in-medium $NN$ elastic
scattering cross section.

\end{abstract}

\pacs{25.70.-z, 21.65.Ef}

\maketitle

\section{INTRODUCTION}

Knowledge on the density dependence of the symmetry energy is
crucial to understand the structure of exotic nuclei, dynamics of
heavy-ion collisions, and many important issues in nuclear
astrophysics such as neutron star cooling and supernova explosive
\cite{liba2008,VBaran2005,JM.Lattimer2004,Steiner2005}. While
great progress has been made to constrain the symmetry energy at
low densities \cite{betty1,betty2,betty3,zhang13,lee13}, the
high-density behavior of the symmetry energy is divergent widely
from interpreting FOPI data
\cite{guo2014,XiaoZG:2009,FengZQ:2010,Russotto2011,xie2013,cozma2013}.
$\pi^-/\pi^+$ ratio was found to be a sensitive probe to the
high-density behavior of the symmetry energy by several transport
models \cite{Yong06,LIba02,Gaitanos04,LiQF05,gao13}. In fact for
pion production, at lower beam energies effects of symmetry energy
and in-medium effect may both become larger
\cite{Zhangfang12,xu13}. It is thus necessary to do a comparative
study of effects of the in-medium $NN$ cross section and the
effects of the symmetry energy on pion production in heavy-ion
collisions at lower beam energies. And these experimental studies
will become possible at facilities that offer fast radioactive
beams, such as NSCL and FRIB in the US, FAIR in Germany, or RIBF
in Japan.

\section{The IBUU Model}

In this study, we adopt the semi-classical transport model IBUU04,
in which the isospin-dependent initial neutron and proton density
distributions of the projectile and target are obtained by using
the Skyrme-Hartree-Fock with Skyrme $M^{*}$ (SM) force parameters
\cite{Friedrich86}. And an isospin- and momentum-dependent
mean-field single nucleon potential is also used, i.e.,
\begin{eqnarray}
U(\rho,\delta,\textbf{p},\tau)&=&A_u(x)\frac{\rho_{\tau'}}{\rho_0}+A_l(x)\frac{\rho_{\tau}}{\rho_0}\nonumber\\
& &+B(\frac{\rho}{\rho_0})^{\sigma}(1-x\delta^2)-8x\tau\frac{B}{\sigma+1}\frac{\rho^{\sigma-1}}{\rho_0^\sigma}\delta\rho_{\tau'}\nonumber\\
& &+\frac{2C_{\tau,\tau}}{\rho_0}\int
d^3\,\textbf{p}'\frac{f_\tau(\textbf{r},\textbf{p}')}{1+(\textbf{p}-\textbf{p}')^2/\Lambda^2}\nonumber\\
& &+\frac{2C_{\tau,\tau'}}{\rho_0}\int
d^3\,\textbf{p}'\frac{f_{\tau'}(\textbf{r},\textbf{p}')}{1+(\textbf{p}-\textbf{p}')^2/\Lambda^2},
\label{buupotential}
\end{eqnarray}
where $\tau=1/2(-1/2)$ for neutrons(protons),
$\delta=(\rho_n-\rho_p)/(\rho_n+\rho_p)$ is the isospin asymmetry,
and $\rho_n$, $\rho_p$ denote neutron and proton densities,
respectively. The parameters $A_u(x)$, $A_l(x)$, $B$,
$C_{\tau,\tau}$, $C_{\tau,\tau'}$ $\sigma$, and $\Lambda$ are all
given in Ref. \cite{Das03}. $f_{\tau}(\textbf{r},\textbf{p})$ is
the phase-space distribution function at coordinate \textbf{r} and
momentum \textbf{p}. Different $x$ parameters can be used to mimic
different forms of the symmetry energy. In this model, reaction
channels on pion production and absorption are
\begin{eqnarray}
NN&\rightarrow& NN,\nonumber\\
NR&\rightarrow& NR,\nonumber\\
NN&\leftrightarrow& NR,\nonumber\\
R&\leftrightarrow& N\pi,
\end{eqnarray}
where $R$ represents $\Delta$ or $N^*$ resonances. The
experimental free-space nucleon-nucleon ($NN$) scattering cross
section and the in-medium $NN$ cross section can be used
optionally. For the later, we use the isospin-dependent in-medium
$NN$ elastic cross section, which is from the scaling model
according to nucleon effective masses
\cite{Negele81,Pan92,Sammarruca13,liba05}
\begin{eqnarray}
R_{medium}(\rho,\delta,\vec{p})&=& \sigma^{medium}_{NN_{elastic}}/\sigma^{free}_{NN_{elastic}}\nonumber\\
&=&(\mu^{*}_{NN}/\mu_{NN})^2,
\end{eqnarray}
where $\mu_{NN}$ and $\mu^*_{NN}$ are the reduced masses of the
colliding nucleon pair in free space and medium, respectively. And
the effective mass of nucleon in isospin asymmetric nuclear matter
is given by
\begin{eqnarray}
m^*_{\tau}=\{1+\frac{m_{\tau}}{p}\frac{dU_{\tau}}{dp}\}^{-1}m_{\tau}.
\end{eqnarray}
From the definition and Eq.~(1), it is seen that the effective
mass depends not only on density and asymmetry of nuclear matter,
but also the momentum of nucleon \cite{Yong10}. For the inelastic
cross section we use the experimental data from free space $NN$
collision since the in-medium inelastic $NN$ cross section is
still very controversial. The total and differential cross
sections for all other particles are taken either from
experimental data or obtained by using the detailed balance
formula. The isospin dependent phase-space distribution functions
of the particles involved are solved by using the test-particle
method numerically. The isospin-dependence of Pauli blockings for
fermions is also considered.

\section{Results and discussions}

\begin{figure} [t]
\resizebox{0.99\columnwidth}{!}{\includegraphics{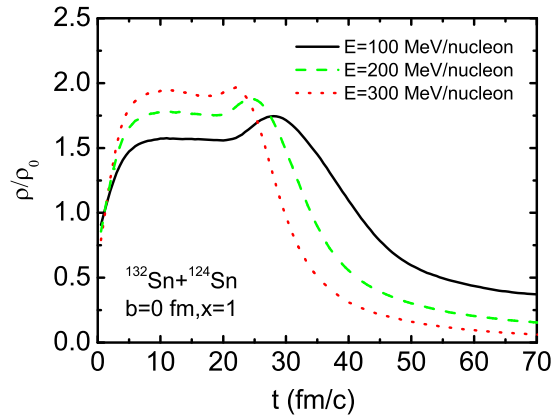}}
\caption{\label{density}(Color online) Maximal baryon densities
reached in central reaction $^{132}Sn+^{124}Sn$ at $E_{beam}$=100,
200 and 300 MeV/nucleon.}
\end{figure}
As shown in Fig.~\ref{density}, the maximal baryon densities
reached in $^{132}Sn+^{124}Sn$ collisions are about 1.5$\sim$2
times saturation density at $E_{beam}$=100, 200 and 300
MeV/nucleon. We can also see that the maximal baryon density
reached increase with incident beam energy. However, existing time
of supradensity matter becomes shorter with increase of beam
energy.

\begin{figure} [t]
\resizebox{0.99\columnwidth}{!}{\includegraphics{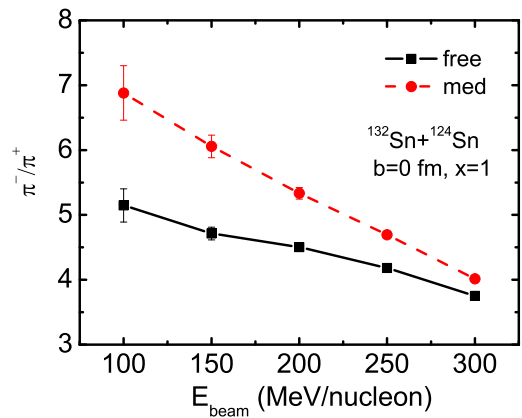}}
\caption{\label{ratiobeam}(Color online) Effects of in-medium $NN$
elastic scattering cross section on the $\pi^-/\pi^+$ ratio in
central collision $^{132}Sn+^{124}Sn$ at $E_{beam}$=100, 150, 200,
250, and 300 MeV/nucleon, respectively.}
\end{figure}
Figure~\ref{ratiobeam} shows effects of in-medium $NN$ elastic
scattering cross section on the $\pi^-/\pi^+$ ratio in central
collision $^{132}Sn+^{124}Sn$ at $E_{beam}$=100, 150, 200, 250,
and 300 MeV/nucleon, respectively. From Fig.~\ref{ratiobeam} we
can see that the value of $\pi^-/\pi^+$ ratio decreases with
increase of beam energy, which is consistent with that in Ref.
\cite{Zhangfang12}. This is partially because the production of
pions is from repeated nucleon-nucleon collisions at higher beam
energies, i.e., a neutron converts a proton by producing $\pi^-$
and subsequent collisions of that proton can convert again to
neutron by producing $\pi^+$. More interestingly, one can see that
the effects of in-medium $NN$ elastic cross section on the value
of $\pi^-/\pi^+$ ratio become larger and larger with decrease of
beam energy. At lower beam energy 100 MeV/nucleon, effects of
in-medium $NN$ elastic cross section on the value of $\pi^-/\pi^+$
ratio can reach about 40\%. When one changes $NN$ elastic cross
section, total $NN$ collision number would also changes
accordingly. As there is certain probability of inelastic process
in total $NN$ collision, $NN$ inelastic collision is thus affected
by $NN$ elastic cross section. At high beam energy, $NN$ inelastic
processes may be more than elastic processes. So elastic $NN$
cross section should have small effects on pion production. But at
low beam energy, pion production is via many $NN$ scatterings,
large number of $NN$ elastic scatterings increases the whole $NN$
scatterings, since there is certain probability of inelastic
process in $NN$ collision, pion production should be affected by
$NN$ elastic scatterings.

\begin{figure} [t]
\resizebox{0.99\columnwidth}{!}{\includegraphics{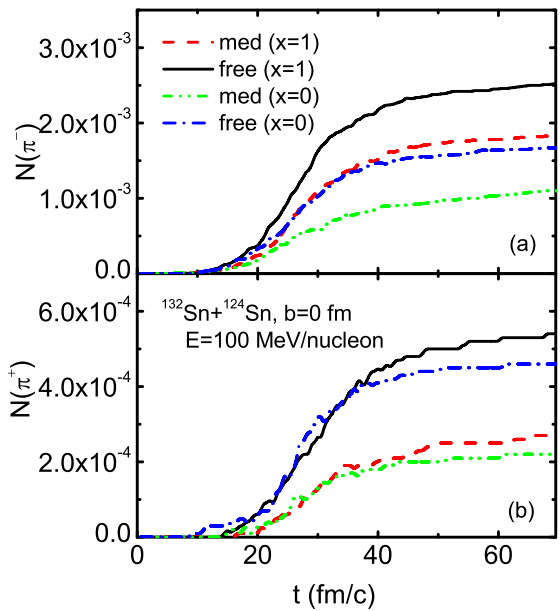}}
\caption{\label{pion-t1}(Color online) Time evolution of $\pi^-$
and $\pi^+$ mesons with different symmetry energies (x=0, x=1) and
different $NN$ elastic scattering cross sections (in-medium, free)
in the central collision $^{132}Sn+^{124}Sn$ at the beam energy of
100 MeV/nucleon.}
\end{figure}
\begin{figure} [htb]
\resizebox{0.99\columnwidth}{!}{\includegraphics{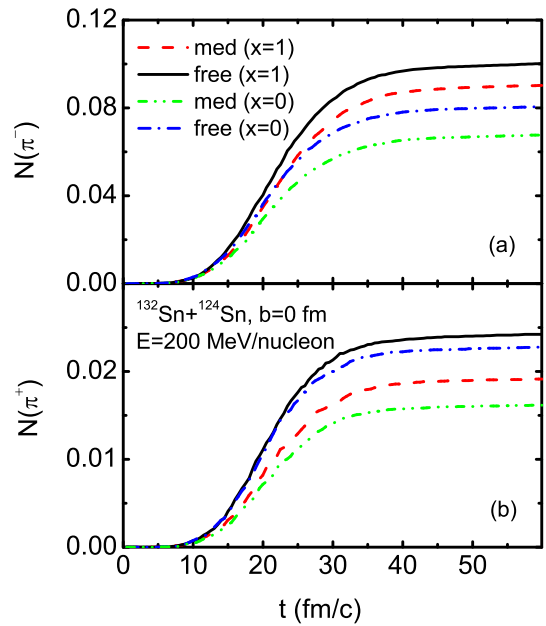}}
\caption{\label{pion-t2}(Color online) Same as
figure~\ref{pion-t1} but at the beam energies of 200 MeV/nucleon.}
\end{figure}
To see more clearly how the in-medium $NN$ cross section affect
$\pi^-/\pi^+$ ratio and effect of the symmetry energy on pion
production, we plot evolution of pion meson production with
different $NN$ cross sections and symmetry energies. Shown in
figure~\ref{pion-t1} and figure~\ref{pion-t2} are time evolutions
of $\pi^-$ and $\pi^+$ mesons with different symmetry energies
(x=0, x=1) and different $NN$ elastic scattering cross sections
(in-medium, free) in the central collision $^{132}Sn+^{124}Sn$ at
beam energies of 100 and 200 MeV/nucleon, respectively. In both
Fig.~\ref{pion-t1} and Fig.~\ref{pion-t2}, we can see that the
effects of in-medium $NN$ elastic scattering cross section on the
$\pi^+$ production is larger than that of $\pi^-$ production.
However, the effects of symmetry energy on the $\pi^+$ production
is smaller than that of $\pi^-$ production. The reason of
in-medium effects on the $\pi^+$ production is larger than that of
$\pi^-$ production is that the reduction factor $R_{medium}$ of
$pp$ (proton-proton) pair is smaller than that of $nn$
(neutron-neutron) pair \cite{Yong10} as well as $pp$ $(nn)$
collision mainly produce $\pi^+$ ($\pi^-$). And due to Coulomb
actions among protons, $\pi^+$ production is also less sensitive
to the symmetry energy. In figure~\ref{pion-t1} and
figure~\ref{pion-t2}, we can also see that for $\pi^-$ production,
effects of in-medium $NN$ cross section are smaller than that of
symmetry energy whereas for $\pi^+$ production, effects of
in-medium $NN$ cross section are obviously larger than that of
symmetry energy. And with increase of beam energy, effects of
in-medium $NN$ cross section become smaller.

\begin{figure} [t]
\resizebox{0.99\columnwidth}{!}{\includegraphics{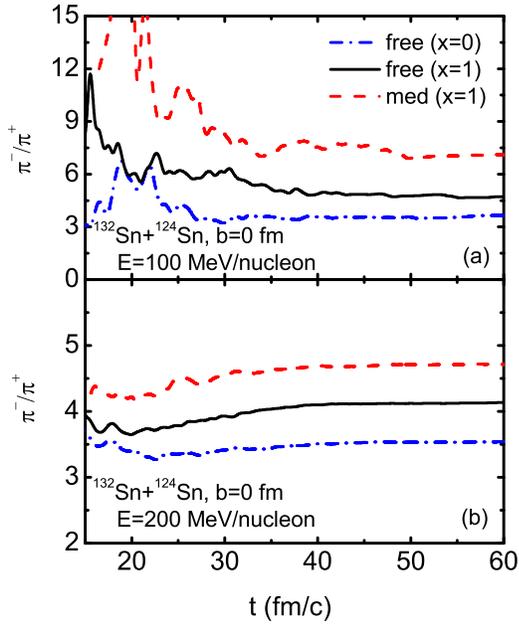}}
\caption{\label{pionratio-t}(Color online) Effects of in-medium
$NN$ cross section and symmetry energy on $\pi^-/\pi^+$ ratio in
central collision $^{132}Sn+^{124}Sn$ at beam energies of 100 and
200 MeV/nucleon, respectively.}
\end{figure}
To see more clearly effects of in-medium $NN$ elastic scattering
cross section and symmetry energy on $\pi^-/\pi^+$ ratio, in
Fig.~\ref{pionratio-t}, we show time evolution of $\pi^-/\pi^+$
ratio with different symmetry energies and $NN$ elastic scattering
cross sections at beam energies of 100 and 200 MeV/nucleon,
respectively. We can clearly see that effects of in-medium $NN$
cross section on the value of $\pi^-/\pi^+$ ratio are about 2
times larger than that of the symmetry energy at incident beam
energy of 100 MeV/nucleon.  However, at incident beam energy of
200 MeV/nucleon, effects of in-medium $NN$ cross section on the
value of $\pi^-/\pi^+$ ratio are almost equal to that of the
symmetry energy.

\begin{figure} [t]
\resizebox{0.99\columnwidth}{!}{\includegraphics{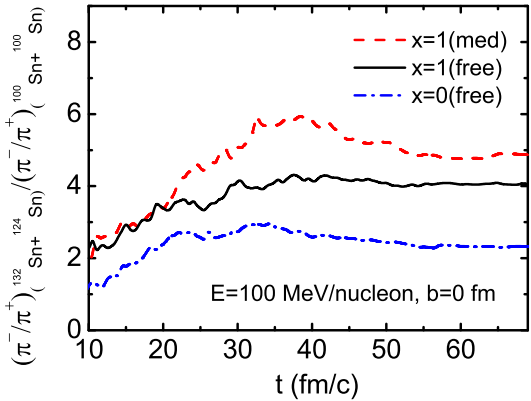}}
\caption{\label{doubleratio-t}(Color online) Effects of medium and
symmetry energy on the double $\pi^-/\pi^+$ ratio from
$^{132}Sn+^{124}Sn$ and $^{100}Sn+^{100}Sn$ at the beam energy of
100 MeV/nucleon.}
\end{figure}
\begin{figure} [t]
\resizebox{0.99\columnwidth}{!}{\includegraphics{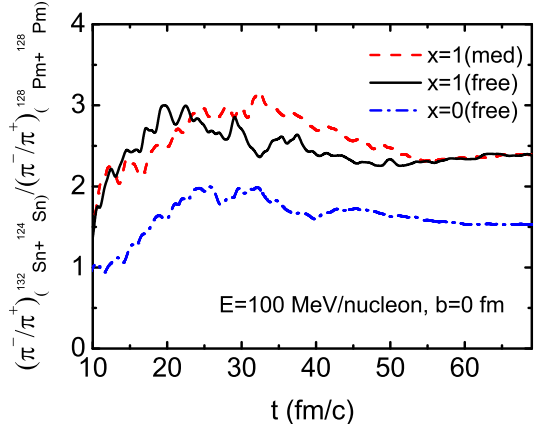}}
\caption{\label{doubleratio-t1}(Color online) Effects of medium
and symmetry energy on the double $\pi^-/\pi^+$ ratio from
$^{132}Sn+^{124}Sn$ and $^{128}Pm+^{128}Pm$ at the beam energy of
100 MeV/nucleon.}
\end{figure}
In order to reduce the effects of in-medium and retain effects of
the symmetry symmetry on $\pi^-/\pi^+$ ratio, we calculate double
ratio of $\pi^-/\pi^+$ using two reaction systems of neutron rich
and neutron poor of same isotopes \cite{Yong06}. Shown in
Fig.~\ref{doubleratio-t} is time evolution of the double
$\pi^-/\pi^+$ ratio from $^{132}Sn+^{124}Sn$ and
$^{100}Sn+^{100}Sn$. It is seen that effects of the in-medium $NN$
cross section on the double $\pi^-/\pi^+$ ratio is about 30\%, but
effects of the symmetry energy is about 42\%. However, from the
top panel of Fig.~\ref{pionratio-t}, we see that effects of the
in-medium $NN$ cross section on the $\pi^-/\pi^+$ ratio is about
50\%, but effects of the symmetry energy is about 25\%. Thus the
double ratio of $\pi^-/\pi^+$ from two reaction systems of neutron
rich and neutron poor of same isotopes can indeed reduce
uncertainties from in-medium properties of hadrons and keep the
effects of symmetry energy. While from Fig.~\ref{doubleratio-t},
we can see that even we use the double ratio of observable,
effects of in-medium $NN$ cross section are still very obvious.
Besides theoretical efforts \cite{Sammarruca13} and advanced
method such as photon emission in heavy-ion collisions
\cite{yongphoto}, we also simulated the double ratio of
$\pi^{-}/\pi^{+}$ from two reaction systems (with the same mass
number of system but different isotopes) $^{132}Sn+^{124}Sn$ and
$^{128}Pm+^{128}Pm$. Interestingly, from Fig.~\ref{doubleratio-t1}
one sees that effects of symmetry energy are kept but the effects
of in-medium $NN$ elastic scattering cross section are almost
fully cancelled out. This is because the latter two neutron-rich
and neutron-poor reaction systems have the same baryon number,
evolutions and distributions of baryon density in the two
reactions are almost the same.

\section{Conclusions}
In summary, we studied effects of in-medium $NN$ cross section and
effects of symmetry energy on $\pi^-/\pi^+$ ratio at lower
incident beam energies. We find that at lower incident beam
energy, for $\pi^-/\pi^+$ ratio, effects of in-medium $NN$ cross
section are larger than that of symmetry energy. The double ratio
of $\pi^-/\pi^+$ from reaction systems of neutron rich and neutron
poor of same isotopes can not fully cancel out the effects of
in-medium $NN$ cross section. However, the double ratio of
$\pi^-/\pi^+$ from reaction systems of neutron rich and neutron
poor of different isotopes but same mass number of reaction system
almost fully cancels out the effects of in-medium $NN$ cross
section.

\section*{Acknowledgments}

The authors are grateful to the C3S2 computing center in Huzhou
Teachers College for calculation support. This work is supported
by the National Natural Science Foundation of China (11375239,
11175219, 11435014), the 973 Program of China (No. 2007CB815004),
the Knowledge Innovation Project(KJCX2-EW-N01) of Chinese Academy
of Sciences.


\begin{thebibliography}{99}
\bibitem{liba2008}B.A. Li, L.W. Chen, and C.M. Ko, Phys. Rep. \textbf{464}, 113(2008).
\bibitem{VBaran2005}V. Baran, M. Colonna, V. Greco, M. Di Toro, Phys. Rep. \textbf{410}, 335(2005).
\bibitem{JM.Lattimer2004}J.M. Lattimer, M. Prakash, Science \textbf{304}, 536(2004).
\bibitem{Steiner2005}A.W. Steiner, M. Prakash, J.M. Lattimer, et al., Phys. Rep. \textbf{411}, 325(2005).

\bibitem{betty1}M.B. Tsang, Yingxun Zhang, P. Danielewicz, M. Famiano, Zhuxia Li, W.G. Lynch, A.W. Steiner, Phys. Rev. Lett. \textbf{102}, 122701 (2009).
\bibitem{betty2}M.B. Tsang, Z. Chajecki, D. Coupland, P. Danielewicz, F. Famiano, R. Hodges, M.Kilburn, F. Lu, W.G. Lynch, J. Winkelbauer, M. Youngs, YingXun Zhang, Prog. Part. Nucl. Phys. \textbf{66}, 400 (2011).
\bibitem{betty3}M. B. Tsang, J. R. Stone, F. Camera, P. Danielewicz, S. Gandolfi, K. Hebeler, C. J. Horowitz, Jenny Lee, W. G. Lynch, Z. Kohley, R. Lemmon, P. Moller, T. Murakami, S. Riordan, X. Roca-Maza, F. Sammarruca, A. W. Steiner, I. Vida$\tilde{n}$a, S. J. Yennello, Phys. Rev. C \textbf{86}, 015803 (2012).

\bibitem{zhang13}Y.X. Zhang, Z. Li, K. Zhao, H. Liu, M. Tsang, Nuclear Science and Techniques \textbf{24}, 050503 (2013).
\bibitem{lee13}K.S. Jeong, S.H. Lee, Nuclear Science and Techniques \textbf{24}, 050506 (2013).
\bibitem{guo2014}W.M. Guo, G.C. Yong, Y.J. Wang, Q. Li, H.F. Zhang, W. Zuo, arXiv:1404.7217 (2014).
\bibitem{XiaoZG:2009}Z.G. Xiao, B.A. Li, L.W. Chen, G.C. Yong, M. Zhang, Phys. Rev. Lett. \textbf{102}, 062502 (2009).
\bibitem{FengZQ:2010}Z.Q. Feng, G.M. Jin, Phys. Lett. B \textbf{683}, 140 (2010).
\bibitem{Russotto2011}P. Russotto, P.Z. Wu, M. Zoric, M. Chartier, Y. Leifels, R.C. Lemmon, Q. Li, J. \L ukasik, A. Pagano, P. Paw\l owski, W. Trautmann, Phys. Lett. B \textbf{697}, 471 (2011).
\bibitem{xie2013}W.J. Xie, F.S. Zhang, Nuclear Science and Techniques \textbf{24}, 050502 (2013).
\bibitem{cozma2013}M.D. Cozma, Y. Leifels and W. Trautmann, Q. Li, P. Russotto, Phys. Rev. C \textbf{88}, 044912 (2013).
\bibitem{Yong06}G.C. Yong, B.A. Li, L.W. Chen and W. Zuo, Phys. Rev. C \textbf{73}, 034603 (2006).
\bibitem{LIba02}B.A. Li, Phys. Rev. Lett. \textbf{88}, 5296 (2002).
\bibitem{Gaitanos04}T. Gaitanos, M. Di Toro, S. Typel, et al., Nucl. Phys. A  \textbf{732}, 24 (2004).
\bibitem{LiQF05}Q.F. Li, Z.X. Li, S. Soff, M. Bleicher, and H. St\"{o}cker, Phys. Rev. C \textbf{72}, 034613 (2005).
\bibitem{gao13}Y. Gao, G.C. Yong, Y.J. Wang, Q.F. Li, W. Zuo , Phys. Rev. C \textbf{88}, 057601 (2013).
\bibitem{Zhangfang12}F. Zhang, Y. Liu, G.C. Yong, and W. Zuo, Chin. Phys. Lett. \textbf{29}, 052502 (2012).
\bibitem{xu13}J. Xu, L.W. Chen, C.M. Ko, B.A. Li, Y.G. Ma, Phys. Rev. C \textbf{87}, 067601 (2013).
\bibitem{Friedrich86}J. Friedrich and P.G. Reinhard, Phys. Rev. C \textbf{33}, 335 (1986).
\bibitem{Das03}C.B. Das et al., Phys. Rev. C \textbf{67}, 034611 (2003); B.A. Li et al., Nucl. Phys. A \textbf{735}, 563 (2004).
\bibitem{Negele81}J. W. Negele and K. Yazaki, Phys. Rev. Lett. \textbf{47}, 71 (1981).
\bibitem{Pan92}V.R. Pandharipande and S. C. Pieper, Phys. Rev. C \textbf{45}, 791 (1992).
\bibitem{Sammarruca13}F. Sammarruca, Eur. Phys. J. {\bf A50}, 22 (2014).
\bibitem{liba05}B.A. Li and L. W. Chen, Phys. Rev. C \textbf{72}, 064611 (2005).
\bibitem{Yong10}G.C. Yong, Eur. Phys. J. {\bf A46}, 399 (2010).
\bibitem{yongphoto}G.C. Yong, W. Zuo, X.C. Zhang, Phys. Lett. B \textbf{705}, 240 (2011).


\end{thebibliography}
\end{document}